\documentclass[a4paper]{article}

\usepackage{amsmath}
\usepackage{amssymb}
\usepackage{graphicx}
\usepackage{natbib}
\usepackage{wrapfig}

\newcommand{\eqlabel}[1]{\label{eq:#1}} % eqn label
\newcommand{\eq}[1]{(\ref{eq:#1})}      % ref to n eqn
\newcommand{\Fig}[1]{Figure~\ref{fig:#1}} % ref to figure
\newcommand{\fig}[1]{Figure~\ref{fig:#1}} % ref to figure
\newcommand{\figlabel}[1]{\label{fig:#1}}
\newcommand{\seclabel}[1]{\label{sec:#1}}
\renewcommand{\sec}[1]{Section~\ref{sec:#1}} % ref to section

\newcommand{\nn}{\nonumber}

\newcommand{\eg}[1]%                    % exempli gratia
  {{\it e.g.\/}\ifx#1.\else\expandafter#1\fi}
\newcommand{\dblfigure}[4]%             % double coulumn figure
  {\begin{figure*}[#1]\centerline{\includegraphics{#2}}\caption[]{#3}\figlabel{#4}\end{figure*}}

\newcommand{\Complex}{\mathbb{C}}       % complex set
\newcommand{\const}{\mathrm{const}}     % constant
\renewcommand{\d}{{\mathrm d}}          % ordinary differential
\newcommand{\df}[2]{\dfrac{\d{#1}}{\d{#2}}}     % ordinary derivative
\newcommand{\hot}{\mathrm{h.o.t.}}      % higher order terms
\newcommand{\mx}[1]{\mathbf{#1}}        % mx and col vec vars
\newcommand{\Real}{\mathbb{R}}          % real set

\newcommand{\Ampl}{A}                   % solution amplitude for visualization
\newcommand{\arcder}{\mathcal{D}_s}     % arclentgh differentiation operator
\newcommand{\B}{{\vec B}}               % filament binormal vector
\newcommand{\bzoeps}{\varepsilon}       % oregonator epsilon
\newcommand{\bzof}{f}                   % oregonator f
\newcommand{\bzoq}{q}                   % oregonator q
\newcommand{\cgla}{\alpha}              % linear dispersion in CGLE
\newcommand{\cglb}{\beta}               % nonlinear dispersion in CGLE
\newcommand{\curv}{\kappa}              % filament curvature
\newcommand{\D}{\mx{D}}                 % diffusion matrix
\newcommand{\Dfac}{D}                   % relative diffusivity
\newcommand{\Du}{D_u}                   % u diffusivity
\newcommand{\Dv}{D_v}                   % v diffusivity
\newcommand{\eps}{\epsilon}             % perturbation small parameter
\newcommand{\FSphase}{\Phi}             % Frenet-Serret phase
\newcommand{\f}{\mx{f}}                 % reaction rates
\newcommand{\Gphase}{\phi}              % geodesic (Fermi) phase
\newcommand{\h}{\mx{h}}                 % generic perturbation
\newcommand{\N}{{\vec N}}               % filament principal normal vector
\renewcommand{\O}{\mathcal{O}}          % the approximation oder 
\renewcommand{\P}{\mx{P}}               % componentwise diffusivity matrix
\newcommand{\R}{{\vec R}}               % R vector (s.w. centre coords) 
\renewcommand{\r}{{\vec r}}             % any pt on the plane or in space
\newcommand{\RF}[1]{\mx{W_{#1}}}        % spatial response function
\newcommand{\T}{{\vec T}}               % filament tangent vector
\newcommand{\U}{\mx{U}}                 % unperturbed soln (s.w.)
\renewcommand{\u}{\mx{u}}               % concentration field
\newcommand{\paralp}{\alpha}            % parameter alpha = epsilon in FHN
\newcommand{\parbet}{\beta}             % parameter beta in FHN
\newcommand{\pargam}{\gamma}            % parameter gamma in FHN
\newcommand{\para}{a}                   % parameter a in Barkley
\newcommand{\parb}{b}                   % parameter b in Barkley
\newcommand{\parc}{c}                   % parameter c=epsilon in Barkley
\newcommand{\tors}{\tau}                % filament torsion
\newcommand{\uc}{u_*}                   % tip definition constant
\newcommand{\vc}{v_*}                   % tip definition constant
\renewcommand{\a}{a}
\renewcommand{\b}{b}
\renewcommand{\c}{c}
\newcommand{\dd}{d}
\newcommand{\e}{e}
\newcommand{\gam}{\gamma}

\newcommand{\w}{w}

\newcommand{\tensc}{\gam_1}
\newcommand{\tenps}{\gam_2}
\newcommand{\rigsc}{\e_1}

\begin{document}
\title{Dynamics of filaments of scroll waves\label{chap:scrolls}}
\author{Vadim N. Biktashev and Irina V. Biktasheva}
\date{Submitted 2014/01/15 as a chapter for \\ ``Engineering of Chemical
  Complexity II''}
\maketitle

\tableofcontents

\section{A brief history and motivation}\seclabel{history}

One of notable events in the history of creation of the new
science of ``cybernetics'' was Norbert Wiener's visit to Arturo
Rosenblueth in Mexico, which resulted in their joint
paper~\citep{Wiener-Rosenblueth-1946}, describing the first
mathematical model of propagation of excitation pulses through a
two-dimensional (2D) continuum, such as heart muscle. An important
assertion of that theory was the possibility of the pulses to
circulate around inexcitable obstacles, with important implications
for understanding certain cardiac
arrhythmias. \citet{Balakhovsky-1965} realized that that circulation
of waves does not in fact require an obstacle, and the excitation wave
may circulate ``around itself'', i.e. turning around its own
refractory tail. Subsequently, such regimes of propagation became
known as ``reverberators'', ``rotors'', ``autowave vortices'' and,
mostly, ``spiral waves'' (see~\fig{spiscr}(a)). With the state of
cardiac electrophysiology at the time, this concept remained a purely
theoretical abstraction until the periodical chemical reaction
  discovered by~\citet{Belousov-1959} came to light and was further
  developed and investigated
  by~\citet{Zhabotinsky-1964-B,Zhabotinsky-1964-PAS}
(Belousov-Zhabotinsky reaction, or just BZ)
 to fruition.
The reaction was
spontaneously oscillating, however in a non-stirred reactor, the
fronts of the reaction oxidation stage were propagating similarly to
electric pulses in cardiac muscle, and the spiral waves made by such
propagation were observed~\citep{Zhabotinsky-Zaikin-1971}.  The
analogy with cardiac excitability was made even more succinct
by~\citet{Winfree-1972} who modified BZ recipe to make the
reaction dynamics excitable rather than oscillatory, and who was the
first to present the BZ reaction and spiral waves to the Western
audience. Being a physical model of cardiac tissue was arguably the
most important use for the BZ reaction, motivating its study for all
these years.

Since then, spiral waves have been observed in a wide variety of
biological, chemical and physical systems, both artificial and in
nature. We mention here just one example, the waves of cAMP signalling
during the aggregation stage of social amoebae \textit{Dictyostelium
  discoideum} ~\citep{Alcantara-Monk-1974,Tyson-Murray-1989}. There
the spiral waves serve as organising centers, as they provide signals
to the individual amoebae where to crawl, to gather and merge into a
multi-cellular organism and thus continue their peculiar lifecycle.

\dblfigure{tbp}{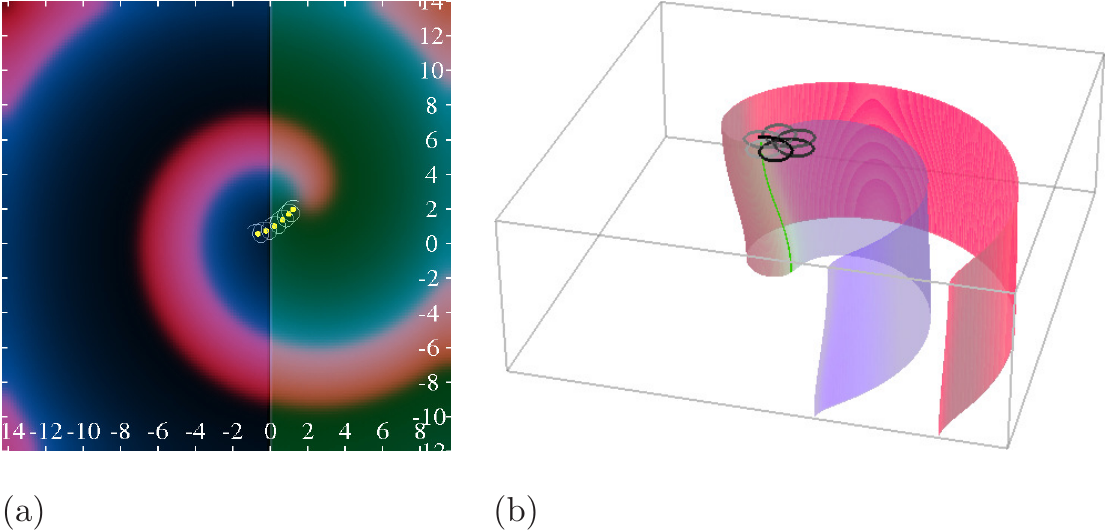}{%
  (a) Snapshot of spiral wave in the Barkley model~\eq{Barkley} ($u$:
  red colour component, $v$: blue colour component), drifting in a
  stepwise inhomogeneity of paramer $\parc$ (green colour component),
  at base parameter values.  The thin white line is the trace of the
  tip of the spiral, defined by $u(x,y,t)=\uc$, $v(x,y,t)=\vc$, in the
  course of a few preceding rotations. Yellow circles are positions of
  the centres calculated as period-averaged positions of the
  tip~\citep{Biktasheva-etal-2010}.
  (b) Snapshot of scroll wave with buckled filament with negative
  tension in a thin layer of medium described by Barkley
  model~\eq{Barkley} \citep{Dierckx-etal-2012}. Shown
  is the surface $u(x,y,z,t)=\uc$, coloured depending on $v$; the
  green line is the instant filament defined as the locus of
  $u(x,y,z,t)=\uc$, $v(x,y,z,t)=\vc$. The grayscale line on the top
  face is the trace of the upper end of the filament.
  Parameter values: in (a), $\para=0.7$, $\parb=0.01$, $\parc=0.025$,
  $\delta\parc=0.001$,$\Du=1$, $\Dv=0$, $\uc=0.5$, $\vc=0.34$, box
  size $24\times24$; in (b), $\para=1.1$, $\parb=0.19$, $\parc=0.02$,
  $\Du=1$, $\Dv=0.1$, $\uc=0.5$, $\vc=0.36$, box size
  $20\times20\times6.9$.
}{spiscr}

Unlike the Wiener and Rosenblueth's 2D theoretical abstraction, real
excitable and oscillatory media, including BZ reaction and heart
muscle, are three-dimensional (3D). Explicit experimental data on
3D extensions of spiral waves were first presented by
\citet{Winfree-1973} in his variant of BZ reaction. He also coined
the term ``scroll waves'' (see~\fig{spiscr}(b)). While
spiral waves rotate around their ``cores'', which can be considered
point-like geometric objects, scroll waves rotate around
``filaments'', line-like geometric objects.  Winfree also called them
``organizing centres''~\citep{Winfree-Strogatz-1983-1}, but in a sense
different from what it means for the social amoebae, as there 
are no living bodies to receive
the signals in the BZ reaction. Rather, the filaments
organize the waves in the sense that the
wavefield in the whole volume follows what happens around the
filament, and the rest of the wavefield can be more or less predicted
using Huygens principle, as shown by Wiener and Rosenblueth.

As spiral and scroll waves do not
require any obstacles to rotate about, they can be located anywhere
within the reactive medium. An inevitable, even if not immediately
evident, consequence of that is that their position can change in
time, i.e. they can drift, as illustrated by~\fig{spiscr}.
Experimental and numerical studies of the drift
have revealed that in many cases it is convenient to consider
spiral waves as ``particles'', interacting with each other or reacting
to external perturbations as localized, point-wise objects, the
location being at the core around which the spiral rotates.  The
scroll waves in 3D have more degrees of freedom:
their filaments can not only move in space, but also change shape. The
phase of rotation may vary along the filament, the feature known as
``twist'' of the scroll wave.  Twist of a scroll wave and curvature of
its filament are specifically
three-dimensional factors of its dynamics.

It also turned out that not only it is \emph{convenient to describe}
motion of spirals and scrolls in terms of their cores and filaments,
but it is \emph{possible to predict} their motion in terms of cores
and filaments coordinates \emph{only} (particularly considering the phase as one of
the coordinates). In this review, we aim to briefly discuss
why this approach works and what sort of results it can produce. The
literature on scroll dynamics is vast and the available space enforce
us to restrict to selected examples in the narrow topic defined by the title of this
chapter. We shall neglect plenty of other interesting results,
including most of twist-related effects and everything related to
competition and meander.

\section{Wave-particle duality of spiral waves}\seclabel{duality}

The possibility to replace consideration of spiral waves by
``particles'', interacting with each other or reacting to external
perturbations, is in a seeming
contradiction with the wave-nature of these objects. The spiral waves
``look'' like non-localized objects, filling up all available space,
but ``behave'' like localized objects. The mathematical nature of this
paradox was brought to the forefront by
\citet{Biktasheva-Biktashev-2003} in terms of perturbative dynamics of
spiral waves, i.e. drift of spirals in response to symmetry-breaking
perturbations, such as spatial gradient of medium parameters or their
resonant periodic modulation in
time. Following~\citep{Biktashev-Holden-1995}, consider a reaction diffusion
system
\begin{equation}\eqlabel{RDS}
\partial_t\u =\f(\u)+\D \nabla^2\u +\eps\h , 
\end{equation}
where
$\u =\u(\r,t)$, $\f,\h\in\Real^{\ell}$, $\D\in\Real^{\ell\times\ell}$,
$\ell\geq2$, $\r\in\Real^2$, 
with perturbation
\(\eps\h =\eps\h(\r,t,\u,\nabla\u,\dots)\), \(||\eps\h|| \ll 1\),
and assume existence, at \(\eps=0\), of
stationary rotating spiral solutions
\begin{equation}\eqlabel{Spiral}
\u(\r,t)
  =\U(\rho(\r-\R),\theta(\r-\R)+\omega t-\FSphase)
\end{equation}
where \(\rho(),\theta()\) are polar coordinates centered at $\R$,
constant $\omega$ is the
angular velocity of spiral wave rotation, which up to the sign is
uniquely defined by medium properties ($\f()$ and $\D$)\footnote{
  See, however, end of \sec{turb} below.
}, and arbitrary constants
$\R=(X,Y)$ and $\FSphase$ are the location of the core of
the spiral and its fiducial (initial) phase at $t=0$.
Then the first order perturbation theory in \(\eps\) gives
solutions close to \eq{Spiral} with $R=X+iY$ and $\FSphase$ slowly varying
according to motion equations
\begin{equation}\eqlabel{GenericDrift}
  \dot{R}=\eps H_1(\R,\FSphase,t),
  \qquad
  \dot\FSphase=\eps H_0 (\R,\FSphase,t) .
\end{equation}
%
% In particular, for the drift under resonant forcing in a spatially
% inhomogeneous medium, this leads to~\citep{Biktashev-Holden-1994-JTB}
% %
% \begin{equation}\eqlabel{InhResDrift}
%   \dot{R} = C(X,Y) + v(X,Y) e^{i\Theta}, 
%   \qquad 
%   \dot{\Theta}=\Omega(t)-\tilde\omega(X,Y),
% \end{equation}
% %
% where \(\Theta\) is the phase difference between the spiral and the
% resonant forcing, \(\Omega\) is the forcing frequency, and
% \(\tilde\omega(X,Y)\) is the own (perturbed) spiral angular frequency
% possibly depending on the current spiral location.
%
The velocities of spatial
drift, \(H_1\),  and temporal/phase drift, \(H_0\),  are linear
functionals of the perturbation,
\begin{equation}\eqlabel{Forces}
H_n=\oint_{t-\pi/\omega}^{t+\pi/\omega} \frac{\omega\,\d{t'}}{2\pi}
  \iint\limits_{\Real^2} \d^2\r  \;
  e^{in(\FSphase-\omega{t'})} \left\langle
    \RF{n}(
      \rho,
      \theta +\omega{t'}-\FSphase)
    ),\h 
  \right\rangle ,
\end{equation}
where 
\(\h\) is evaluated at the unperturbed solution \eq{Spiral}.

\dblfigure{tbp}{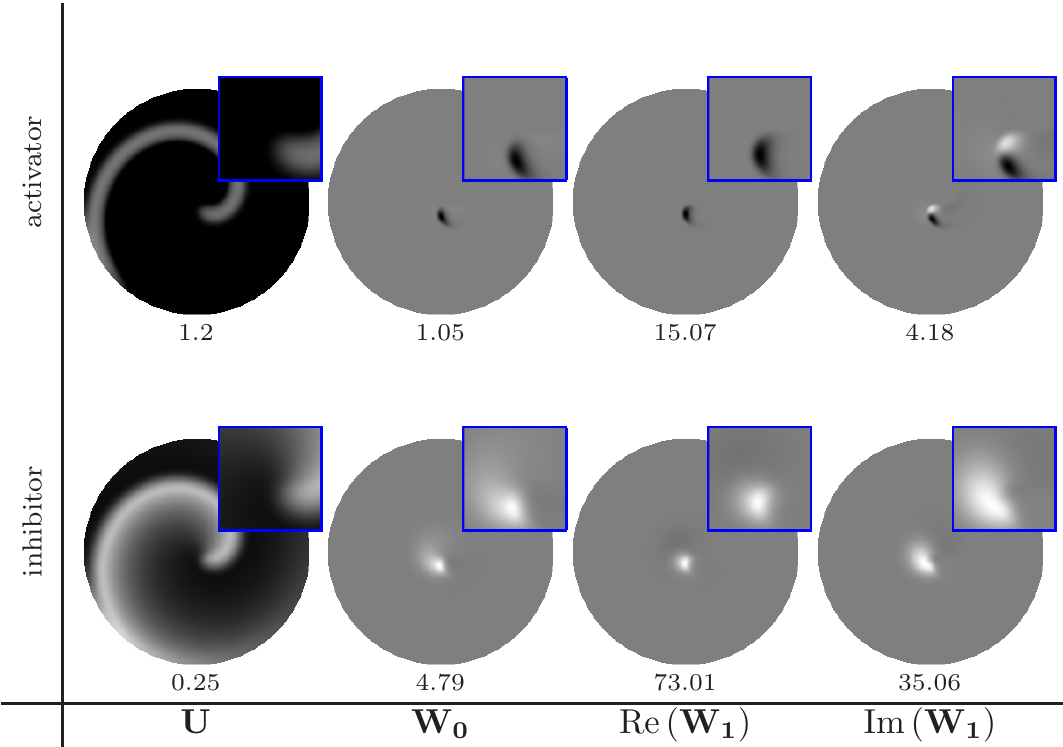}{%
  Density plots of the components of the spiral wave solution and its
  response functions in the Oregonator model~\eq{Oregonator} for
  $\bzoeps=0.06$, $\bzof=1.75$, $\bzoq=0.002$, $\Du=1$, $\Dv=0.6$, the
  radius of the disk is 15 dimensionless units.  In each plot, white
  corresponds to a value $\Ampl$ and black corresponds to $-\Ampl$
  where $\Ampl$ is chosen individually for each plot, \eg\ for the
  activator component of $\U$, $\Ampl=1.2$. The grey periphery of the
  plots in columns 2--4 corresponds to 0.  The central areas are also
  shown magnified in the small corner panels. %
}{bzorf}

\dblfigure{tbp}{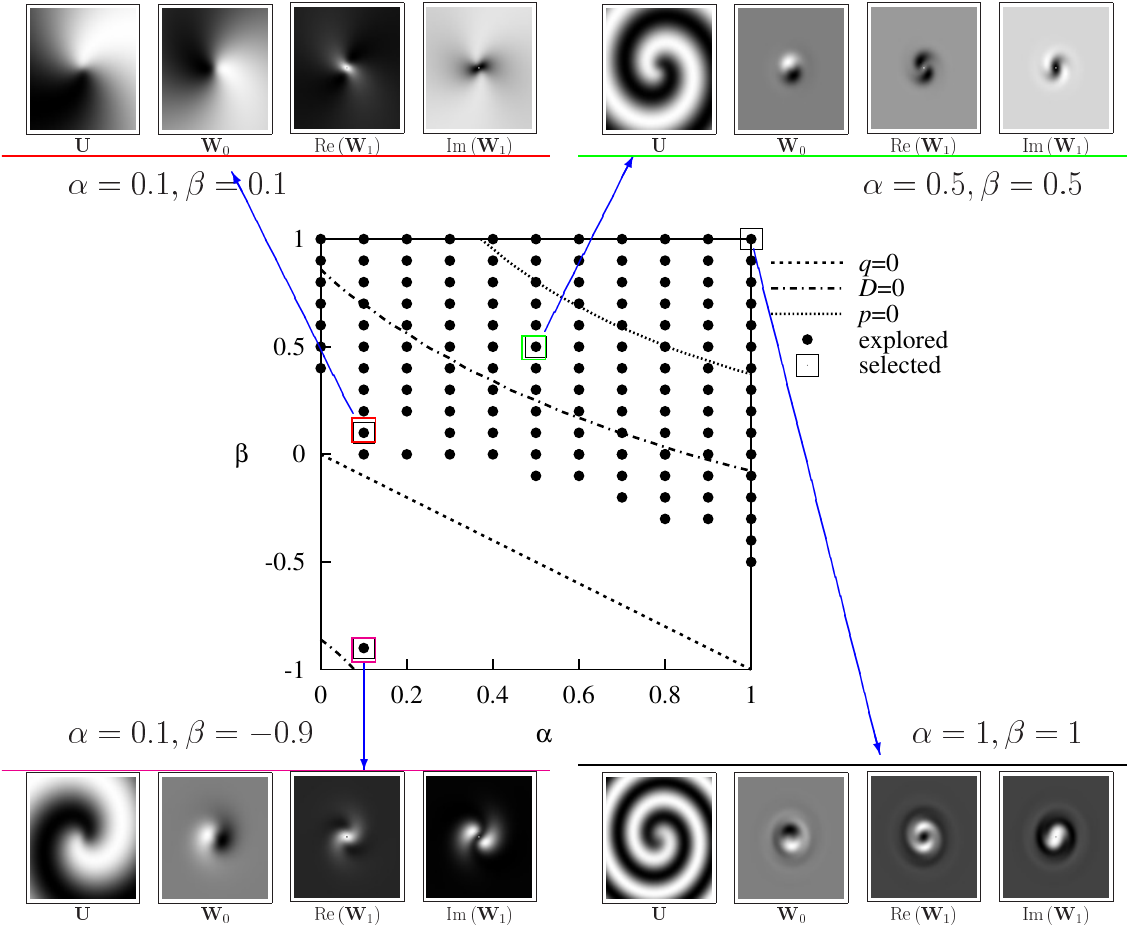}{%
  Density plots of the components of the spiral wave solution and its response functions
  in the complex Ginzburg-Landau equation~\eq{CGLE}, at various
  parameter values. In the legend: 
  $p = 2\left(
    \alpha\beta - 1 + 
    k^2\left(
        3 + 2\beta^2 - \alpha\beta
    \right)
  \right)/(1+\beta^2)$, 
  $q = - 4k\left(
    (\alpha+\beta)(1-k^2)
  \right)(1+\beta^2)$, 
  $D=(p/3)^3+(q/2)^2$, %
  and $k=k(\alpha,\beta)$ is the asymptotic wavenumber of the
  spiral~\citep{Biktasheva-Biktashev-2001}. %
}{cglrf}

The kernels \(\RF{0,1}\) of these functionals, which we call
response functions (RFs), are critical eigenfunctions of the adjoint
linearized problem. These eigenfunctions are dual to the
eigenfunctions of the linearized problem produced by the generators of
the Euclidean symmetry group, sometimes called Goldstone Modes (GMs). The
``wave-particle' duality then reduces to the difference between these
eigenfunctions. The GMs, constructed from spatial derivatives of the
spiral wave solution, are non-localized. The RFs,
however, are essentially localized, i.e. exponentially decay far from
the core of the spiral. This is of course only possible because the
linearized problem is not self-adjoint.

The spiral waves have localized response functions in both excitable
and oscillatory media.  \Fig{bzorf} shows response functions in the
two-component Oregonator~\citep{Tyson-Fife-1980} model of the BZ reaction,
\begin{eqnarray} \eqlabel{Oregonator}
  \partial_t u &=& \frac{1}{\bzoeps}\left( u(1-u) - \bzof v
    \frac{u-\bzoq}{u+\bzoq}\right) + \Du \nabla^2 u, 
  \nn\\
  \partial_t v &=& u - v + \Dv \nabla^2 b , 
\end{eqnarray}
for a choice of parameters $\bzoeps,\bzof,\bzoq$ that gives
excitable dynamics.
\Fig{cglrf} shows response functions in the
complex Ginzburg-Landau equation (CGLE)~\citep{Kuramoto-Tsuzuki-1975},
\begin{equation} \eqlabel{CGLE}
  \partial_t w = w - (1 - i\cgla) w|w|^2 + (1+i\cglb) \nabla^2w,
  \qquad
  w = u + i v \in \Complex, 
\end{equation}
which is the ``archetypical'' oscillatory reaction-diffusion
model, in the sense that it is a normal form of a reaction-diffusion system near a
supercritical Hopf bifurcation in its reaction
part.
In CGLE, the response functions 
are localized for all sets of parameters where 
stable spiral wave solutions exist, with qualitative changes
across critical lines in the parameter
space~\citep{Biktasheva-Biktashev-2001}. Notice the non-monotonous
behaviour (``halo'') in RFs close to the Eckhaus instability line
(\fig{cglrf}, bottom right inset). This can have phenomenological
implication for the dynamics of spirals and scrolls,
discussed later.

Apparently, the defining condition for the RFs' localization is the
direction of the group velocity: a spiral wave will have localized
RFs and behave as a localized object if and only if it is a source of
waves, so that far from the core, the group velocity is directed
outwards~\citep{Biktashev-etal-1994,Sandstede-Scheel-2004}.

\dblfigure{tbp}{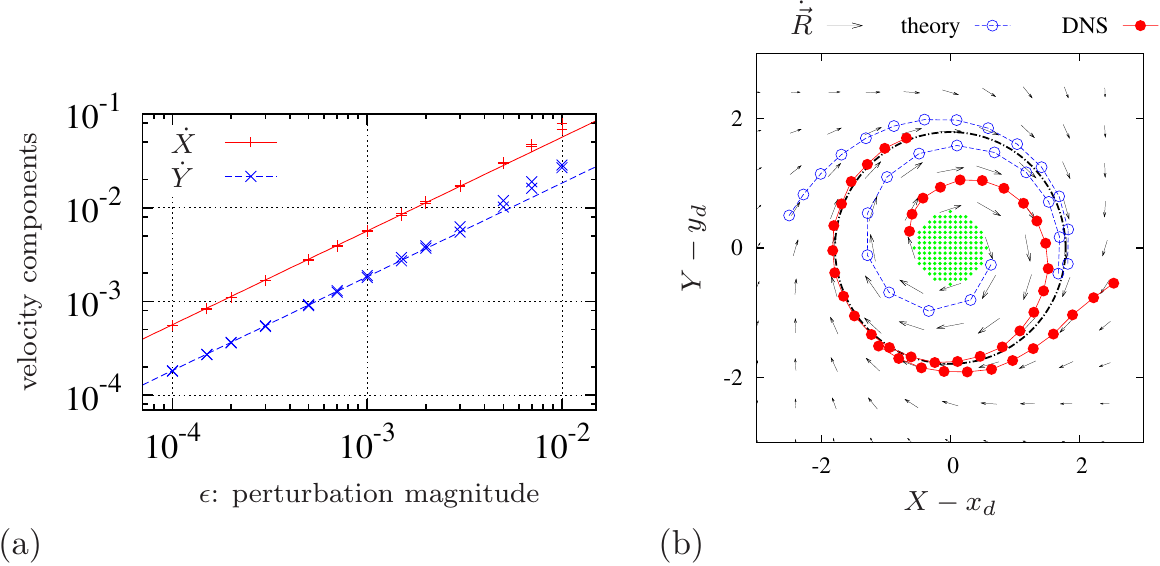}{%
  Drift of spiral waves: asymptotic theory vs direct numerical
  simulations. (a) The velocity of the drift caused by the gradient of
  parameter $\para$ in the Barkley model~\eq{Barkley}, symbols for DNS
  and lines for the asymptotics, for base values
  of parameters $\para=0.7$, $\parb=0.01$, $\parc=0.025$, $\Du=1$,
  $\Dv=0$. (b) The trajectories of the drift caused by a 
  disk-shaped inhomogeneity in parameter $\pargam$ in the
  FitzHugh-Nagumo model~\eq{FHN}, at base values $\paralp=0.3$,
  $\parbet=0.68$, $\pargam=0.5$, $\Du=1$, $\Dv=0$. The small
  arrows indicate drift velocities as predicted by the asymptotic
  theory, and the filled and open circles show the instantaneous centers
  of rotation of the spiral wave, measured with the interval of one
  period of
  rotation. See~\citep{Biktashev-etal-2010,Biktasheva-etal-2010} for
  detail.  %
}{drift}

\Fig{drift} reproduces two selected results illustrating how well the
perturbation theory works, for two classical simplified excitable
media models, the FitzHugh-Nagumo model~\citep{Winfree-1991-C}
\begin{eqnarray} \eqlabel{FHN}
  \partial_t u &=& \frac{1}{\paralp}\left( u - \frac{u^3}{3}-v\right) +
  \Du \nabla^2 u, \nn\\
  \partial_t v &=& \paralp \left( u - \parbet v + \pargam \right) +
  \Dv \nabla^2 v,
\end{eqnarray}
and Barkley model~\citep{Barkley-1991}
\begin{eqnarray} \eqlabel{Barkley}
  \partial_t u &=& \frac{1}{\parc}
  u(1-u)\left(u-\frac{v-\parb}{\para}\right) 
  + D_u \nabla^2 u, 
  \nn\\
  \partial_t v &=& u - v 
  + D_v \nabla^2 v .
\end{eqnarray}
The trajectories in \fig{drift}(b) correspond to the case 
of the response function $\RF{1}$ 
with non-monotonic
behaviour (the ``halos''
in~\fig{cglrf}, bottom right, are an extreme case of such
  non-monotonic RFs). Here, small inhomogeneity attracts a
spiral wave at larger distances and repels it at smaller distances.
This alternating attraction/repulsion
causes the spiral wave to ``orbit'' around at a certain stable
distance, where the radial component of the ``interaction force''
vanishes.

There are important examples of spirals dynamics due to factors
that are not small perturbations in the sense of~\eq{RDS}, even though
their action on the spirals is small. These include interaction of
spiral waves with boundaries, and their interaction with
each other (which may be considered as interaction of each spiral with
the boundary between their ``domains of influence''). These
interactions are weak when the distance from the spiral core to the
boundary or between the spiral cores is large.  The mathematical
aspects of particle-like behaviour in such cases are less clear. In
the few examples where analytical answers are known, this seems to be
associated with the exponential growth of solutions of the
non-homogeneous linearized problem with the free term given by the
spatial gradient of the spiral wave solution, see
e.g. \citep{Biktashev-1989}. This is also stipulated by the outward
direction of the group velocity. Therefore both localization
properties seem to be equivalent. That is, if a spiral wave does not
feel a weak inhomogeneity when far from it, it will not feel a
non-flux boundary at the same distance. Although this
equivalence is quite plausible physically, mathematically it is still
an open question.

\section{Perturbative dynamics of scrolls, and tension of
  filaments}\seclabel{tension}

The perturbative dynamics of spiral waves
can be extended to scroll waves.
In 3D, there are interesting phenomena even in absence of any
symmetry breaking perturbations. 
Following~\citet{Keener-1988}, consider 
a generic reaction-diffusion
system \eq{RDS} in $\r\in\Real^3$ at $\h\equiv0$, 
and assume, as before, existence of stationarily
rotating spiral solutions~\eq{Spiral} in $\Real^2$. 
A simple extension of spiral wave solution to the third spatial dimension
is called a straight scroll wave. More generically, 
a scroll wave solution in \(\Real^3\) may be viewed
as a solution of the form
\begin{equation} \eqlabel{curvedscroll}
  \u(\R +\N \rho\cos\theta + \B \rho\sin\theta,t) =  
  \mathbf{U}(\rho,\theta+\omega t-\FSphase) + \mathcal{O}(\eps), 
\end{equation}
where 
\(\eps\)
is now a formal small parameter measuring
deformation of a scroll wave compared to the straight scroll, 
\(\R=\R(\sigma,t)\)
is the parametric equation of filament
position at time 
\(t\),
\(\FSphase=\FSphase(\sigma,t)\) 
is the rotational phase distribution,
\(\N=\N(\sigma,t)\) 
and 
\(\B=\B(\sigma,t)\) 
are the unit principal normal and binormal vectors to the filament at point
\(\R(\sigma,t)\).
Vectors
\(\N\) and
\(\B\)
together with tangent vector
\(\T\)
make a Frenet-Serret triple. In terms of the arclength
differentiation operator
\(
  \arcder{u}(\sigma)=|\partial_{\sigma}{\vec
    R}|^{-1}\partial_{\sigma}u(\sigma)
\),
the tangent vector is 
\(\T=\arcder{\R}\),
the curvature
\(\curv\)
and the normal unit vector
\(\N\) 
are defined by 
\(\curv\N=\arcder{\T}\),
and the binormal vector 
\(\B =\T \times\N\)
completes the triad.
The resulting filament's
equation of motion, at small filament
curvature, \(\curv=\O(\eps)\), and small twist,
\(\arcder\FSphase=\O(\eps)\), can be written as~\citep{Biktashev-etal-1994}
\begin{equation}\eqlabel{filmotion}
\partial_t\R  = \tensc \arcder^2\R  
  + \tenps \left[ \arcder\R  \times  \arcder^2\R  \right]
  + \O(\eps^2) ,
\end{equation}
and it is decoupled from the evolution equation for the phase
$\Phi(\sigma,t)$. Equation \eq{filmotion} is written in the assumption
that parameter $\sigma$ of the filament is chosen in such a way that a
point with a fixed \(\sigma\) moves orthogonally to the filament
(hence no component along $\T$).

The Frenet-Serret description is easy to understand but it has a
significant disadvantage: it becomes degenerate at 
zero filament curvature, \(\curv=0\).  An alternative
description, free from this disadvantage, is in terms of Fermi-Walker
coordinates, corresponding to a Levi-Civita (torsion-free metric)
connection along the filament (hereafter called Fermi coordinates for
brevity). This changes the scroll's phase definition
but does not affect~\eq{filmotion} as it is decoupled. %  from the phase.

The coefficients $\tensc$, $\tenps$ in~\eq{filmotion} are calculated using the response
functions \(\RF{1}\) of the corresponding 2D spiral waves as
\begin{equation}\eqlabel{tension}
 \tensc + i \tenps = -\frac12 \int\limits_0^{\infty}\oint
   \left[\RF{1}(\rho,\theta)\right]^+
   \D 
      e^{-i\theta} \left(
      \partial_{\rho}-\frac{i}{\rho}\partial_{\theta}
    \right) \mathbf{U}(\rho,\theta)
    \;\d\theta
    \;\rho\,\d\rho
   .
\end{equation}

Now let us consider the total length of the filament, defined at each
\(t\):
\begin{equation}\eqlabel{totlength}
S(t)=\int \d{s}
  = \int \left| \partial_{\sigma}\R  \right| \,\d\sigma .
\end{equation}
Differentiation of \eq{totlength}, with account of \eq{filmotion} and using
integration by parts, reveals that, neglecting boundary effects (absent
for closed filaments and vanishing for smooth impermeable boundaries),
the rate of change of the total length is 
\begin{equation}\label{lengthrate}
\df{S}{t} = - \tensc \int \curv^2
  \,\d{s} + \O(\eps^2).
\end{equation}
This implies that, within the
applicability of the perturbation theory, unless the filament is
straight, the total length of the
filament decreases if \(\tensc>0\) and increases if \(\tensc<0\).
Hence the coefficient \(\tensc\) was called ``filament tension''
in~\citep{Biktashev-etal-1994}.

\dblfigure{tbp}{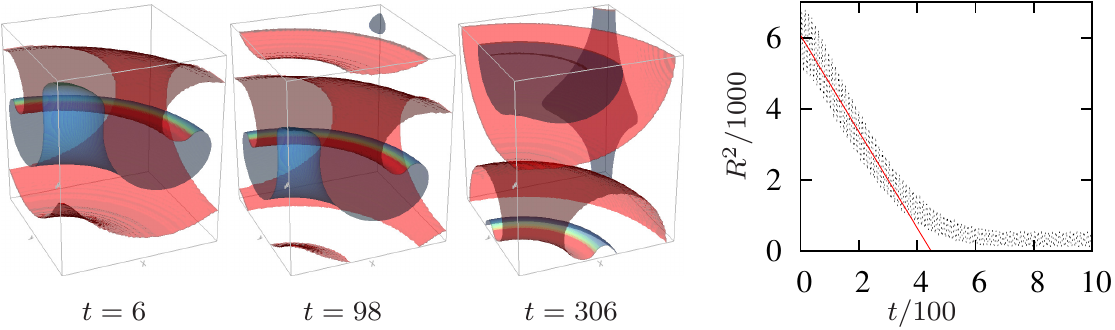}{
  Dynamics of a quarter of a scroll ring in FitzHugh-Nagumo
  model~\eq{FHN}, at $\paralp=0.30$, $\parbet=0.71$, $\pargam=0.5$,
  $\Du=1$, $\Dv=0$ 
  (positive filament tension) in a box $33\times33\times40$ with
  periodic boundary conditions in the $z$ direction. The right panel
  shows evolution of the square of the radius of the instant scroll filament,
  defined as the locus $u=0$. Equation
  \eq{filmotion} predicts that the square of the period-average of the
  radius depends linearly on $t$, with the slope $-2\tensc$. The
  parameters for this example are taken from~\citep{Foulkes-etal-2010}.
}{ring}

Filament tension can be found via the asymptotic rate of shrinkage or
expansion of large scroll rings with exact axial symmetry
(see~\fig{ring}). 
%, in which case the problem becomes mathematically
% equivalent to that of a drift of a spiral wave in an electric field. 
This % correspondence
is only formally valid for scroll rings of
sufficiently large radii in the unbounded space. Depending on the model,
shrinkage of a scroll ring with positive
filament tension may lead to its collapse, or may stop at some finite
radius, while the ring continues to drift along its symmetry
axis~\citep{Brazhnik-etal-1987,Skaggs-etal-1988}, see
\fig{ring}. Possible theoretical explanations of this may involve
higher-order corrections to~\eq{filmotion} and/or interaction of
different pieces of the scroll ring with each other, in the same way
as 2D spirals interact; which if either of these is the dominant
reason in any particular case is, as far as we are aware, currently an
open question. Apart from simple scroll rings, there are more
complicated structures with twisted, linked and/or knotted filaments
that can be persistent in some models,
see~\citep[pp.483--490]{Winfree-2001} for the long story.

Filament tension depends on parameters of the medium, typically
becoming negative in media with lower exitability, where spiral waves
have larger cores \citep{Panfilov-Rudenko-1987}. This can be
substantiated by the ``kinematic'' theory of excitation waves
\citep{Brazhnik-etal-1987}. However, 
% clearly not all excitable media
% satisfy the assumptions of that kinematic theory as 
there are
exceptions to the rule: e.g. 
\citet{Alonso-Panfilov-2008} found an example
where negative filament
tension is observed at high excitability.

A simple but fundamental result is that when diffusivities of all the
reagents are equal, $(\D)_{ij}=\Dfac\delta_{ij}$, then $\tenps=0$ and
$\tensc=\Dfac>0$~\citep{Panfilov-etal-1986}. A less trivial result is
about filament tension in the CGLE~\eq{CGLE},
where it has been shown that $\tenps=0$,
$\tensc=1+\cglb^2>0$, see e.g. \citep{Aranson-Kramer-2002} for
discussion and references. So in both these cases the filament tension
is guaranteed to be positive. We do not know of any generic results
about negative filament tension.

A higher-order asymptotic equation for filament motion 
was obtained by \citet{Dierckx-Verschelde-2013}.
Unlike the leading
order~\eq{filmotion}, it is coupled to the evolution of
the phase: 
\begin{eqnarray} \eqlabel{hiorder}
  \partial_t \Gphase &=& 
  % \om_0+
  \a_0\w^2+\b_0\curv^2+\dd_0\arcder\w + \hot, \nn\\
  \partial_t \R &=& 
  (\gam_1+\a_1\w^2+\b_1\curv^2+\dd_1\arcder\w) \arcder^2\R \nn\\
  && +
  (\gam_2+\a_2\w^2+\b_2\curv^2+\dd_2\arcder\w) \arcder^\R \times \arcder^2\R \nn\\
  && +
  \c_1\w\arcder^3\R+\c_2\w\arcder\R\times\arcder^3\R \nn\\
  && -
  \e_1\arcder^4\R-\e_2\arcder\R\times\arcder^4\R
  + \hot,
\end{eqnarray}
where $\phi$ is the scroll phase measured in the Fermi frame, and $\w$
is the corresponding twist,
$\w=\arcder\Gphase=\arcder\FSphase-\tors$, where $\tors$ is the
filament torsion, $\tors=\B\cdot\arcder\N$.  The vector-function
$\R(s,t)$ in~\eq{hiorder} is the ``virtual filament'' which has to be
defined more precisely than~\eq{curvedscroll},
and the coefficients are defined as integrals
involving spiral wave
response functions, similar to but more complicated than~\eq{tension};
see~\citep{Dierckx-Verschelde-2013} for detail.

\section{Scroll wave turbulence}\seclabel{turb}

\dblfigure{tbp}{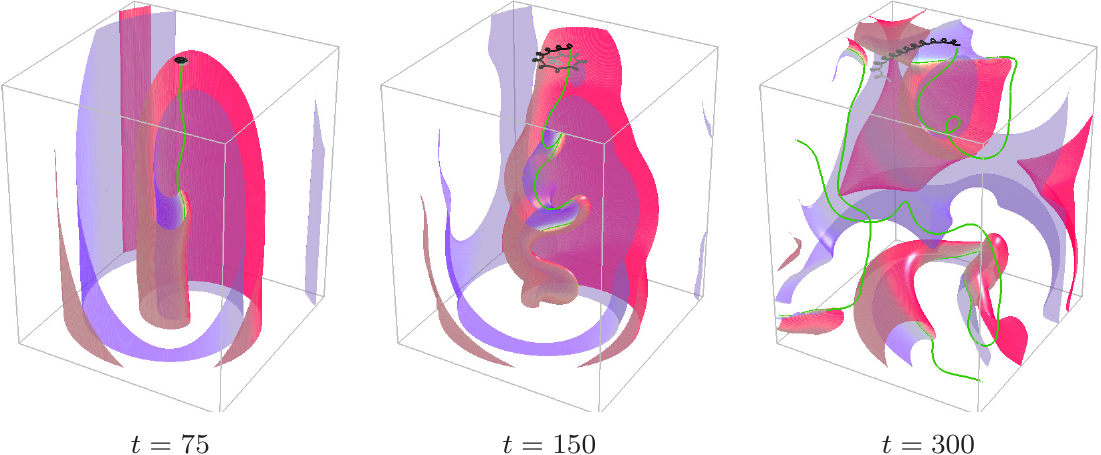}{%
  Negative filament tension instability causes scroll wave
  turbulence. Barkley model \eq{Barkley} with parameters $\para=1.1$,
  $\parb=0.19$, $\parc=0.02$, $\Du=1$, $\Dv=0$, box size
  $40\times40\times50$, instant filaments (green) defined as
  $u=v=0.5$. Wavefronts are cut out by clipping planes halfway through the
  volume, to reveal the filaments~\citep{Dierckx-etal-2012}. %
}{bklturb}

A negative filament tension implies that the straight scroll of a
sufficient lentgh should be unstable.
It was recognized rather early, that unless restabilized by
some mechanism beyond~\eq{filmotion}, such instability can
lead to complex spatio-temporal behaviour, possibly
chaotic~\citep{Brazhnik-etal-1987,Biktashev-etal-1994}.  A particular
interest for this complicated behaviour was due to its possible
relation to cardiac fibrillation, with which it would have a number of
phenomenological features in
common~\citep{Biktashev-etal-1994,Winfree-1994-S,Biktashev-1998}.
The predicted ``scroll wave 
turbulence'' was confirmed by
numerical simulations, first in the FitzHugh-Nagumo model~\eq{FHN}
\citep{Biktashev-1998}, and then in Barkley model~\eq{Barkley}
\citep{Alonso-etal-2004}, Oregonator model of the BZ
reaction~\eq{Oregonator}~\citep{Alonso-etal-2006-JPCA}, and Luo-Rudy
model of heart ventricular tissue~\citep{Alonso-Panfilov-2007} to name
a few; see~\citep{Alonso-etal-2013} for a recent review.

Based on the generic results mentioned above, chances to observe
scroll wave turbulence mediated by negative filament tension in a BZ
reaction should be higher when some of the reagents are
immobilized, since scalar diffusivity matrix implies positive filament
tension; in the excitable rather than oscillatory regime, since
tension in the CGLE is always positive; and preferably in the ``lower
excitability'' case. The latter prediction concurs with
Oregonator simulations by~\citet{Alonso-etal-2006-JPCA}.

\begin{wrapfigure}[18]{R}{0.5\textwidth}
  \vspace*{-0.5\baselineskip}
  {\includegraphics{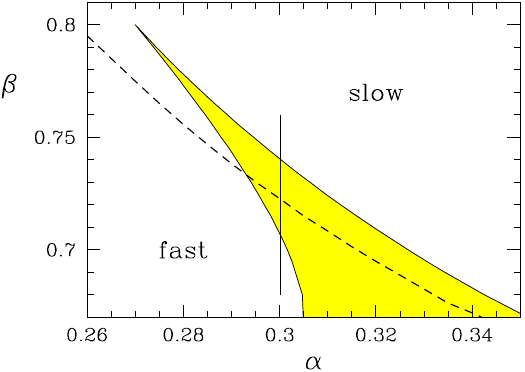}}
  \caption[]{Part of the parameter space in the FitzHugh-Nagumo
    model~\eq{FHN}, giving alternative spiral wave solutions, at fixed
    $\pargam=0.5$.  The dashed line corresponds to zero tension. 
    From~\citet{Foulkes-etal-2010}.}
  \figlabel{fhnalt}
\end{wrapfigure}

We mentioned earlier that the angular velocity $\omega$ of the spiral
rotation is uniquely determined by the properties of the
medium. This is not strictly
true. 
Typically, there is a discrete set of possible $\omega$
values, and in some  cases  there  may  be more then  one of them. 
\citet{Winfree-1991-PD} identified a set of parameters in the
FitzHugh-Nagumo model~\eq{FHN}, which could, depending on initial
conditions, support two alternative spiral wave solutions with
different $\omega$.
\Fig{fhnalt} shows a region in the parameter
space with this property.
Each of the alternative 
spiral waves is stable against small perturbations, but larger
perturbations can convert one sort of spiral to the
other. \citet{Foulkes-etal-2010} investigated such conversions and
some implications for the dynamics of scroll waves in 3D.  One
nontrivial effect observed there is shown in \fig{fhnconv}.

\dblfigure{tbp}{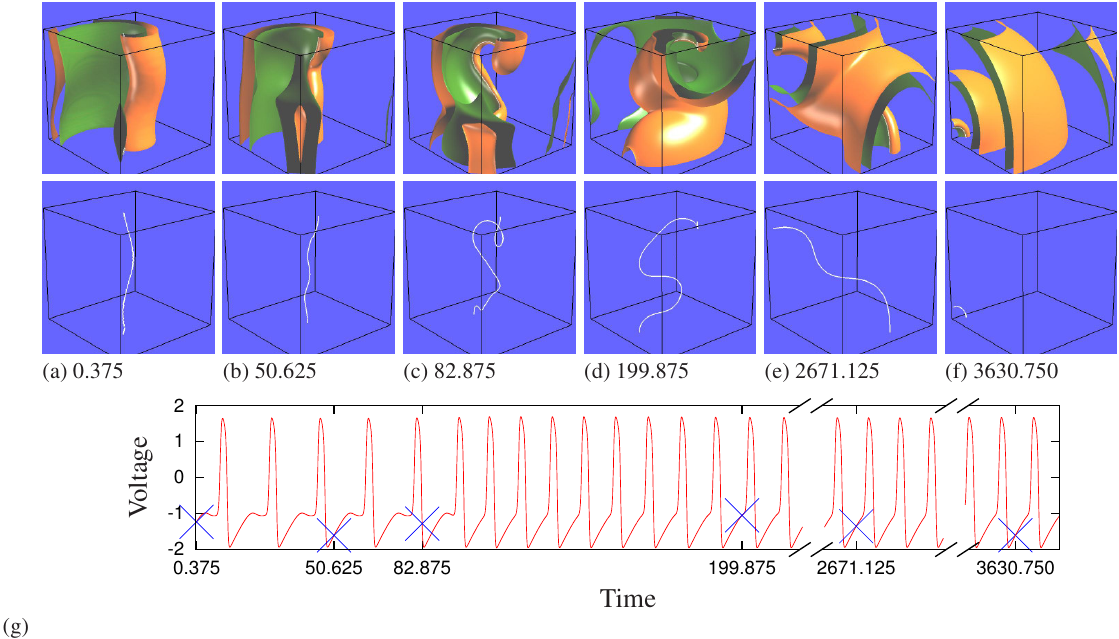}{
  Evolution of a slow helical scroll
  with negative filament tension at 
  $\paralp=0.3$, 
  $\parbet=0.71$, 
  $\pargam=0.5$, 
  $\Du=1$, $\Dv=0$, box size $50\times50\times50$. 
  (a-f) (Top) isosurfaces of the $u$-field; (middle)
  filament ($u=v=0$) only.
  (g) action potentials with blue crosses marking the
  times at which the snapshots were taken. 
  The helix initially expands and turbulizes correspondingly to its negative tension, but then
  converts to its positive-tension alloform and contracts. From~\citet{Foulkes-etal-2010}. 
}{fhnconv}

% \section{Other mechanisms of scroll turbulence}\seclabel{other}

\dblfigure{tbp}{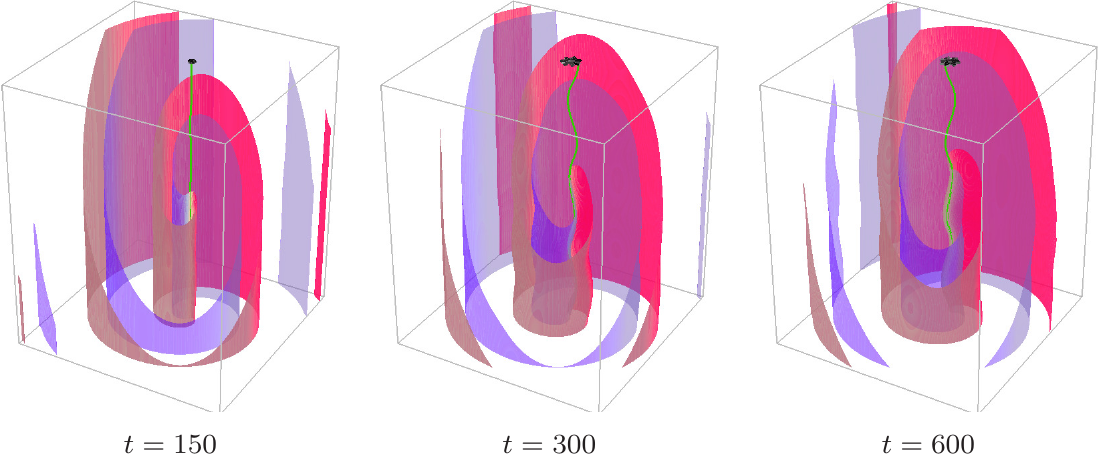}{%
  Short-wave filament instability causes restabilized ``wrinkled''
  filament. Barkley model with parameters $\para=0.66$, $\parb=0.01$,
  $\parc=0.025$, $\Du=1$, $\Dv=0$, box size 
  $40\times40\times50$, instant filaments (green) defined as 
  $u=v=0.5$. Wavefronts are cut out by clipping planes~\citep{Dierckx-etal-2012}. %
}{bklwrink}

% There are other mechanisms that can produce scroll
% turbulence. 
% Negative filament tension implies that a straight scroll is unstable
% with respect to a spectrum of perturbations, that includes those of
% arbitrarily large wavelengths along the scroll axis.  
In various
models, the straight scrolls may become unstable through mechanisms
different from the negative tension
in~\eq{filmotion}. \citet{Henry-Hakim-2002} found that some
parameter changes in Barkley model~\eq{Barkley} can cause
finite-wavelength instability of a scroll, while the filament tension
remains positive. A different type of finite-wavelength
instability at positive filament tension was found in CGLE
by~\citet{Aranson-Bishop-1997}. They
interpreted it in terms of ``self-acceleration''
of spiral waves, at parameter values where the dynamics of
spiral waves in 2D is not well described by the
perturbative dynamics~\eq{GenericDrift} and further corrections are
required. Interestingly, these two finite-wavelength instabilities
have different outcomes: in CGLE, this leads to
turbulent-like behaviour \citep{Aranson-Bishop-1997,Reid-etal-2011} visually similar to that shown
in~\fig{bklturb}, whereas
in Barkley model, it leads to re-stabilized ``wrinkled'' or
``zig-zag'' shape of filaments, as shown
in~\fig{bklwrink}~\citep{Henry-Hakim-2002,Dierckx-etal-2012}. Clearly,
a mere linearized theory can not describe the
outcome of any such instability and more detailed study would be
required to make any predictions there.

Finally, complex turbulent-like behaviour of
scroll waves may be related to spatial heterogeneity. Among these, we
mention the cases described by Fenton and Karma in
 models with spatially varying anisotropy, aimed at representing
characteristic features of cardiac ventricular
muscles,
\begin{equation}\eqlabel{RDSaniso}
  \partial_t\u = \f(\u) +
  \sum\limits_{i,j=1}^3 \partial_i \left( D^{i,j}(\r) \partial_j
  \right) \P \u , 
\end{equation}
where $\P\in\Real^{\ell\times\ell}$ is a constant matrix representing
relative diffusivities of the components
~\citep{Fenton-Karma-1998a,Fenton-Karma-1998b}. Another often
considered possibility is scroll turbulence due to purely
2D mechanisms which make spiral waves
unstable, e.g.~\citep{Panfilov-Holden-1990-PLA} and which of course reveal
themselves in 3D as
well~\citep{Clayton-Holden-2003,Clayton-etal-2005,Clayton-2008,Reid-etal-2011}.

\section{Rigidity of scroll filaments: pinning and buckling}\seclabel{buck}

The scroll wave turbulence mediated by negative filament tension is an
essentially 3D phenomenon: it happens when spiral waves
in a 2D medium with the same parameters are perfectly stable. So
such turbulence is not observed in quasi-two-dimensional domains,
say in thin layers of reactive medium. A gradual
increase of the reactive layer thickness 
reveals that between the stable quasi-two-dimensional
behaviour and the three dimensional turbulence, there are intermediate
regimes, one of which is the ``buckled filament'' illustrated
in~\fig{spiscr}(b) for the Barkley model. Similar regimes were observed in
simulations of Luo-Rudy cardiac model
by~\citet{Alonso-Panfilov-2007}. A qualitative and quantitative
explanation of such filament buckling proposed by
\citet{Dierckx-etal-2012} was based on a simplified version of the
higher-order motion equations~\eq{hiorder}. In particular, it gives
the expression for the critical layer thickness, above which the
buckling
bifurcation happens, in the form $L_*=\pi|\rigsc/\tensc|^{1/2}$, where
$\rigsc$ is the coefficient at the fourth arclength derivative in
\eq{hiorder}, and in this sense it is analogous to the rigidity of an
elastic beam. So, within this mechanical analogy, the stability
threshold for a straight scroll is determined by the
interplay between filament tension $\tensc$ (negative of the
``mechanical stress'') and the filament ``rigidity'' $\rigsc$, which
is similar to the Euler's buckling instability of a beam under a
stress, hence the term ``buckling'' used to describe this deformation
of the scrolls.

\dblfigure{tbp}{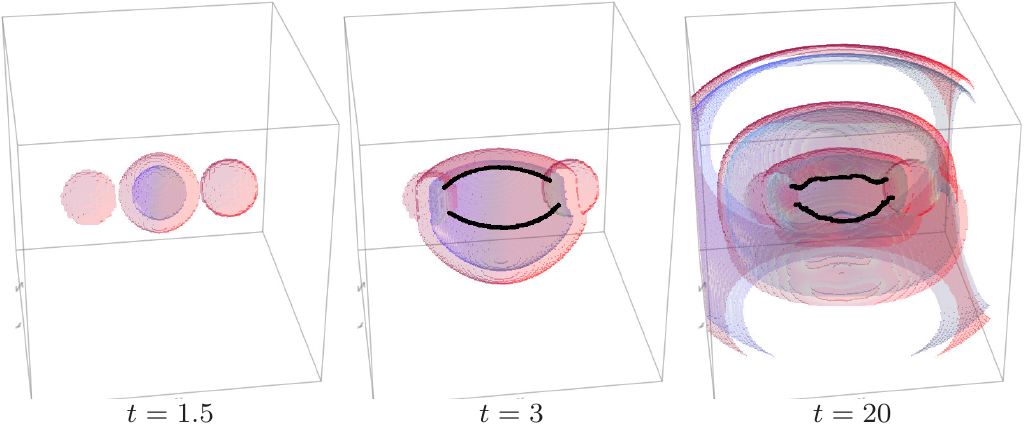}{
  Schematic of the experiments done by
  \citet{Jimenez-Steinbock-2012}. The upper half of a spherical wave
  is cut when it is passing through the beads thus forming two almost
  semi-circular filaments. The filaments become pinned to the beads
  and evolve to a stable configuration which depends on their tension,
  rigidity and interaction force between them. Oregonator
  model~\eq{Oregonator} with parameters 
  $\bzoq=0.002$, 
  $\bzof=1.75$,
  $\bzoeps=0.06$,
  $\Du=\Dv=0.5$, 
% def real umin 0; def real uc 0.25; def real umax 1.2;
% def real vmin 0; def real vc 0.05; def real vmax 0.25;
% r=[u] r0=umin r1=umax
% g=[v] g0=vmin g1=vmax
% 0,0.2			filament field 1, its constant
% 1,0.3			filament field 2, its constant
%  0.2*1.2=0.24
%  0.3*0.25=0.075
  $\uc=0.24$, $\vc=0.075$, 
  box size $50\times50\times50$. 
  Wavefronts are cut out by clipping planes.
}{loop}

Experimental evidence of scroll filament rigidity was
demonstrated in BZ reaction, with filaments pinned to inexcitable
inclusions~\citep{Jimenez-Steinbock-2012,Nakouzi-etal-2014},
see~\fig{loop}. In these experiments, the steady shapes of the
filaments were determined, as well as their relaxation dynamics.
The authors were able to describe the steady
shapes using a variant of~\eq{hiorder},
with added phenomenological
description of the interaction of filaments with each other. Fitting
the theoretical curves to the experimentally found filament shapes
allowed quantitative experimental measurement of the filament
rigidity~\citep{Nakouzi-etal-2014}.

\section{Filament statics, geodesic principle and Snell's law}\seclabel{geodesic}

% The above theoretical concepts on scroll filament dynamics were for
% isotropic and homogeneous media. 
For cardiac applications, 
anisotropy and inhomogeneity are very important. We have already
mentioned above the instability and scroll turbulence related to
inhomogeneous anistoropy. The ``opposite'' of scroll turbulence is the
situation when scroll dynamics converges to a stable equilibrium
position. For positive filament tension, in a spatially uniform, isotropic 
medium and when interaction with boundaries or other filaments can be
neglected, the answer is straightforward: a straight filament,
stretching in any direction.  Motivated by numerical simulations of
scroll waves in models with anisotropic
diffusion such as~\eq{RDSaniso},
and by experiments with re-entrant
excitation waves in cardiac tissue%
% \citep{Berenfeld-Pertsov-1998}?
, \citet{Wellner-etal-2002} came up
with a ``geodesic hypothesis'': that 
for given positions of the filament end points (say if the filament is anchored
to inexcitable inclusions), the steady scroll filament shape will be a geodesic in a metric related to
the diffusivity tensor, $g_{ij}=(D^{-1})_{ij}$. For instance, in a
cardiac muscle, the preferrable orientation of a scroll filament
would be along the fibers, i.e. direction of the maximal diffusivity
of the transmembrane potential. As~\citet{tenTusscher-Panfilov-2004}
noted, the metric defining the steady-state geometries of filaments,
can be conveniently formulated in terms of excitation wave propagation
time: given its end points, the filament will follow the
\emph{quickest} path connecting the end points. This hypothesis 
was confirmed by~\citet{Verschelde-etal-2007} who 
generalized the
equation~\eq{filmotion} for the anisotropic media, under the
assumption of $\det(D^{ij})=\const$. An empirical generalization of
this geodesic principle for non-uniform $\det(D^{ij})$, based on
simulations using Barkley model, was suggested
by~\citet{Wellner-etal-2010}, in the form $g_{ij}=\det(D^{ij})
(D^{-1})_{ij}$. An interesting study has been performed by
\citet{Zemlin-etal-2014}, who investigated in numerical
simulations the analogue of ``Snell's law''for a filament crossing a
boundary between media with different diffusivities, following from
the geodesic principle, and the limits of its applicability.

\bibliographystyle{plainnat}    % Bibliography: Author-Date system
\bibliography{scrolls}      % pls. call your database here

% Restore the standard value of the \columnsep parameter,
% to avoid affecting other chapters' formatting
% \setlength{\columnsep}{\savecolumnsep}

\end{document}